\documentclass[%
 aip,
 amsmath,amssymb,
 reprint,%
]{revtex4-1}

\usepackage{graphicx}
\usepackage{dcolumn}
\usepackage{bm}
\usepackage{xcolor}
\usepackage[utf8]{inputenc}
\usepackage[T1]{fontenc}
\usepackage{mathptmx}
\usepackage{etoolbox}
\usepackage{braket}

\makeatletter
\def\@email#1#2{%
 \endgroup
 \patchcmd{\titleblock@produce}
  {\frontmatter@RRAPformat}
  {\frontmatter@RRAPformat{\produce@RRAP{*#1\href{mailto:#2}{#2}}}\frontmatter@RRAPformat}
  {}{}
}%
\makeatother

\begin{document}


\title{Nanophotonic Quantum Random Number Generator
Using Fiber In-Line Single Photons}

\author{K. Muhammed Shafi}
\email{mkodakkaden@hbku.edu.qa}

\affiliation{
One Photon Systems, C V Raman Avenue, Sadashivanagar,  Bengaluru 560080, India}
\affiliation{
Qatar Center for Quantum Computing, College of Science and Engineering, Hamad Bin Khalifa University, Doha, Qatar}
\author{Samlan Chandran Thodika}
\affiliation{
One Photon Systems, C V Raman Avenue, Sadashivanagar, Bengaluru 560080, India}
\author{Saif Al-Kuwari}
\affiliation{
Qatar Center for Quantum Computing, College of Science and Engineering, Hamad Bin Khalifa University, Doha, Qatar}
\author{Kali P. Nayak}
\affiliation{
Department of Engineering Science, University of Electro-Communications, Tokyo 182-8585, Japan}


\begin{abstract}
We demonstrate a quantum random number generator (QRNG) using single photons from a quantum dot coupled to an optical nanofiber. The key principle of this photonic QRNG is the randomness of single-photon paths arising from equiprobable coupling of spontaneously emitted single photons into two counter-propagating guided modes of the optical fiber. The source exhibits high single-photon purity with  $g^{(2)}(0)=0.12\pm0.01$. The system generates quantum random bits  at rates up to $0.74\pm0.03$ Mbps, exhibits very low  autocorrelation and passes all tests of the NIST Statistical Test Suite. The nanofiber-based photonics platforms hold the potential for a compact, efficient, and intrinsically alignment-free photonic QRNG and can be implemented for potential applications in quantum cryptography and key distribution applications.


\end{abstract}

\maketitle

Random number generation is an essential resource in many applications such as quantum key distribution, cryptography, Monte Carlo simulations, and many others \cite{QKD, Schindler, Monte, Game}.  Classical methods of random number generation, resulting in pseudo-random numbers, often rely on deterministic algorithms, making them potentially predictable and thus vulnerable in high-security applications \cite{PRNG}. Inherent randomness of quantum mechanical systems can create true and unpredictable randomness, and are called quantum random number generator (QRNG) \cite{QRNG1}. Recent progress in controlling quantum systems and extracting quantumness shows a promising replacement for conventional deterministic pseudo-random numbers \cite{QRNGcom1, QRNGRev2017}.

Over the years, numerous QRNG schemes have been developed and refined across various quantum systems. These include implementations using superconducting quantum computers \cite{ QRNG_QC2}, measuring photon number \cite{ PN2011}, quantum phase noise\,\cite{QCL10}, quantum vacuum states \,\cite{QVS10} and phase fluctuations\,\cite{XQM12}. 
Among these platforms, photons are widely regarded as the medium of choice due to their intrinsic quantum randomness with stable room-temperature operations. Their integration with mature photonic technologies enables high-speed, scalable, and practical QRNGs, making them more accessible than other quantum systems.  Subsequently, quantum random bits have been successfully generated by exploiting the randomness in photons’ time of arrival \cite{QRNGP_Time2}, polarisation \cite{BB10}, and phase \cite{QRNG_Phase1}. Additionally, path-based QRNGs typically employ a beam splitter to direct photons along two possible paths probabilistically. Avalanche photo detectors are then used to register the photon detections, producing binary random outcomes. This approach provides a simple and effective prototype for realising QRNGs. For example, the generation of quantum random bits utilising entangled photon pairs and path-entangled single photons from the spontaneous parametric down-conversion process has been successfully demonstrated \cite{ KCK09, ShafiQRNG}. 

The performance of QRNGs primarily hinges on two factors: the bit generation rate and the quality of randomness. Notably, single-photon sources (SPS) have been theoretically shown to yield higher entropy per raw bit, thereby enhancing the statistical robustness of the generated randomness \cite{Entropy2016}. This principle has motivated experimental demonstrations of QRNGs utilizing single photons from diverse quantum emitters, including nitrogen-vacancy centers in diamond \cite{CGW19}, defects in gallium nitride \cite{LCF20}, and in hexagonal boron nitride \cite{WKT20}.  The bit generation rate, on the other hand, is largely governed by the brightness of the SPS. In this context, solid-state quantum emitters, such as quantum dots or atom-like defects, are promising SPS candidates due to their near-unity quantum efficiency and easy integration into nanophotonic systems \cite{Senellart}. 
However, achieving high bit rates also necessitates the efficient collection and delivery of emitted single photons to detection and electronics modules, which remains a critical technical challenge in the practical deployment of SPS-based QRNGs.

To address these limitations, nanophotonic waveguides 
have demonstrated significant advancements by leveraging the strong transverse confinement of photonic modes, surpassing the limitations of free-space optics \cite{Senellart}. 
In particular, tapered optical fibers with sub-wavelength diameter waists, known as optical nanofibers (ONFs), offer a unique fiber in-line platform for efficiently collecting single-photon fluorescence from quantum emitters \cite{famsan1}. A key advantage of this technique lies in the strong transverse confinement of the guided mode in the ONF, which adiabatically evolves into a single-mode fiber (SMF) with near-unity efficiency \cite{Sile1,Kali1}. This seamless transfer of the modes enables an alignment-free, fiber-coupled platform for single-photon collection, bridging nanoscale emitters to macroscale infrastructure while eliminating bulky optics.
Consequently, ONF-based quantum photonic platforms have attracted significant attention, evolving from proof-of-principle lab demonstrations to technological prototypes \cite{ShafiPlasmon,NanoQT}.

In this work, we experimentally demonstrate a path-based QRNG using single photons emitted by a colloidal quantum dot (QD) coupled to an ONF. The QRNG operates by randomly routing single photons between the two output paths of the fiber, with the randomness arising from the symmetric dipole emission pattern and the inherent unpredictability of the spontaneous emission direction. By encoding a binary bit from a single-photon detection event at these two guided modes, we generate a random bit. The device’s quantumness is certified via second-order photon correlation measurement, while the randomness of the generated bit stream is evaluated by measuring the autocorrelation coefficient and by using the National Institute of Standards and Technology (NIST) randomness test suite \cite{RSN01}, ensuring its suitability for cryptographic and other applications.

\begin{figure}[!ht]

\centering

 \includegraphics[width=\columnwidth]{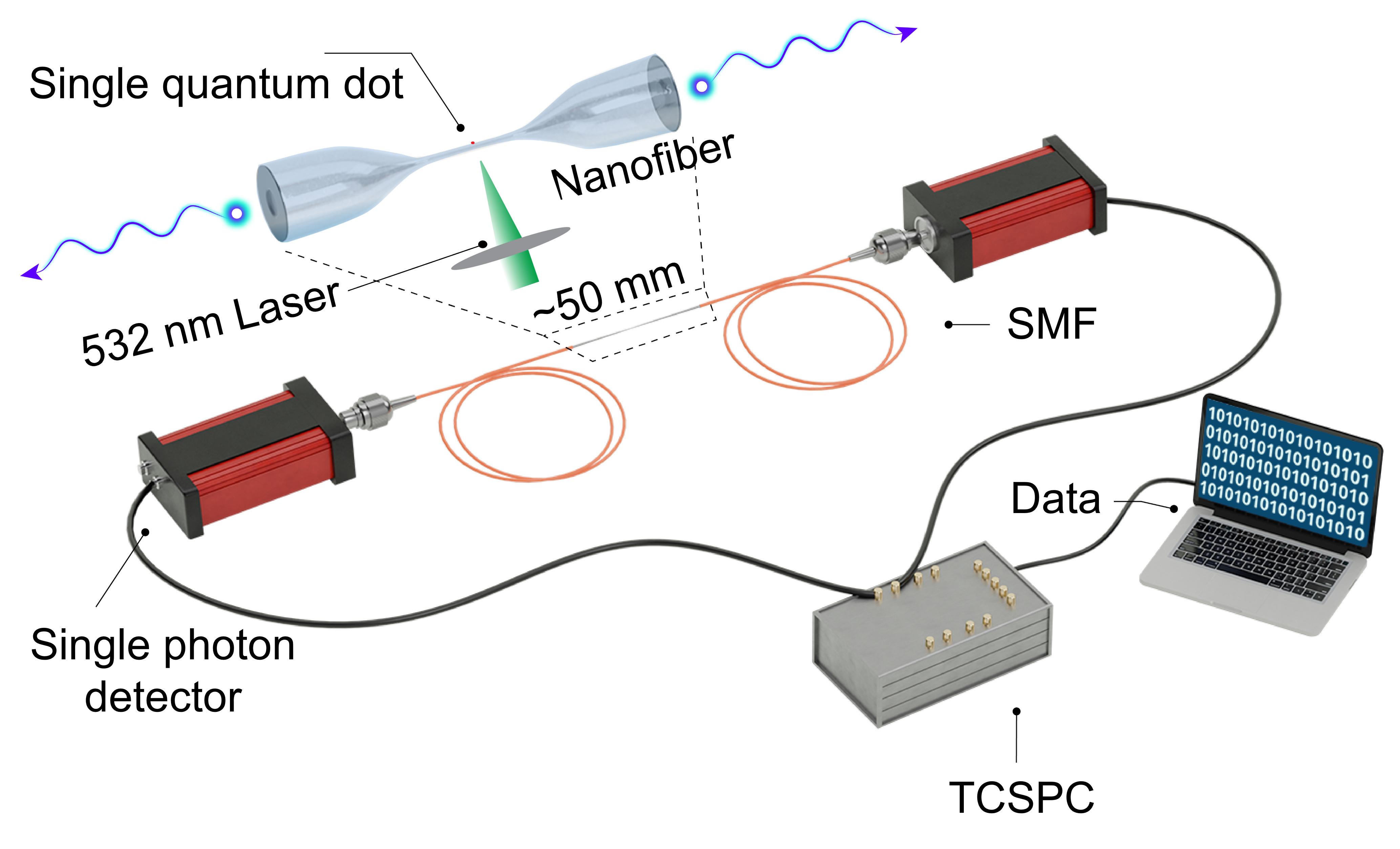}
\caption{
Schematic of the QRNG. The single QD on the ONF  is excited using a 532~nm laser. Single photons emitted from the QD are coupled into the guided modes of the ONF and adiabatically into a SMF. A 560~nm long-pass color-glass filter (not shown) is inserted between the SMF and the detectors to suppress scattered excitation light. The photons propagating along the two fiber paths are detected using a pair of single-photon detectors and time-stamped using a time-correlated single-photon counting module. The single-photon detection events in the two paths are encoded as binary bits to generate one-bit quantum random numbers.
}

\label{Fig1}
\end{figure}

A schematic diagram of the QRNG is shown in Fig. \ref{Fig1}. The experimental setup is based on a single QD deposited on an ONF. The ONF with optical transmission of $\sim$99\% was fabricated using a heat and pull technique by adiabatically tapering a commercial SMF \cite{ONFFab}. The ONF's waist diameter was fixed to 310±10 nm to maximise the coupling efficiency of single photons from the QD into the fiber-guided modes. This enables the coupling of photons at the theoretical coupling efficiency limit of 22$\%$ \cite{famsan1,Yalla1}. The length of ONF waist region was measured to be 2.5 mm, while the adiabatic region was 48 mm long.
The QD used in the experiment was a thick-shell CdSe/ZnS colloidal QD with a size of $\sim$8 nm. These QDs exhibit a room-temperature photoluminescence centred at a wavelength of 640 nm with a spectral width of approximately $\sim$20 nm \cite{Shafi1}. 
The procedure for device preparation is as follows. Single QD, dispersed in toluene, was deposited onto an ONF using a computer-controlled needle dispenser system. The dispenser consists of a tapered glass tube filled with a QD colloidal solution and a tungsten needle.  The solution at the needle tip containing single QDs is then deposited onto the ONF surface by gently bringing the needle tip into contact with the ONF as described in our previous work \cite{ShafiSPS}.  

We first characterized the QRNG by measuring the purity and brightness of the single photons, prior to analyzing the generated bit sequence.  
The QD was excited using a lensed fiber coupled to a continuous-wave laser at a wavelength of 532 nm. The excitation laser was linearly polarized, with its polarization axis oriented perpendicular to the ONF axis. The lensed fiber was mounted on a 3D translation stage, allowing precise adjustment of its tip position relative to the QD, to achieve the smallest possible spot size, which is 10 $\mu$m in the present setup. This enables us to keep the system operational for hours, even with a few microns of mechanical position drift.  To filter photons from the scattered excitation laser, a 560~nm long-pass colour glass filter was inserted between the single-mode fiber and the detectors. All measurements were conducted under ambient laboratory conditions.

The single-photon characteristics of the emitted photons are verified by measuring the second-order photon correlation function, $g^{(2)}(\tau)$, in a Hanbury Brown-Twiss interferometer.  Unlike a conventional setup with the balanced beam splitter, this interferometer is implemented with the photons at the two ends of the fiber. This is possible when the photons couple equiprobably into the guided mode in both directions.
The photons from both sides of the fiber were detected at the two fiber-coupled single-photon detectors ($D_1$ and $D_2$), and the $g^{(2)}(\tau)$ was obtained from the arrival times of the photons recorded using a time-correlated single-photon counter (TCSPC). Here $\tau$ is the time delay between the arrival of photons at the two detectors. The photoluminescence (PL) intensity of the QD at different excitation intensities was measured from the photon-count rate at one detector.

The random-bit generation protocol is as follows. Each photon-detection event at $D_1$ or $D_2$ is independently time-stamped using the TCSPC module with a temporal resolution of 4~ps and assigned the bit value 0 or 1, respectively. The detection events recorded in the two channels are combined and sorted according to their timestamps, thereby generating one random bit per detection event. This path-encoded binary assignment forms the basis of our QRNG scheme. It is worth noting that the equiprobable coupling of single photons into the two counter-propagating guided modes is an intrinsic property of the low-loss optical nanofiber platform. This inherent symmetry naturally yields nearly balanced detection rates at $D_1$ and $D_2$, providing the unbiased binary statistics required for high-quality random bit extraction. Residual imbalance due to practical imperfections, such as detector efficiency mismatch or connector losses, was corrected by introducing additional calibrated losses in the fiber-to-detector coupling, thereby restoring balanced detection probabilities required for unbiased random bit generation. To validate the statistical randomness of the generated data, a 2-Mbit raw random-bit sequence was experimentally generated at an operating rate of 0.66~Mbps and was directly subjected to autocorrelation analysis and the NIST Statistical Test Suite.

\begin{figure}[!ht]

\centering

 \includegraphics[width=\columnwidth]{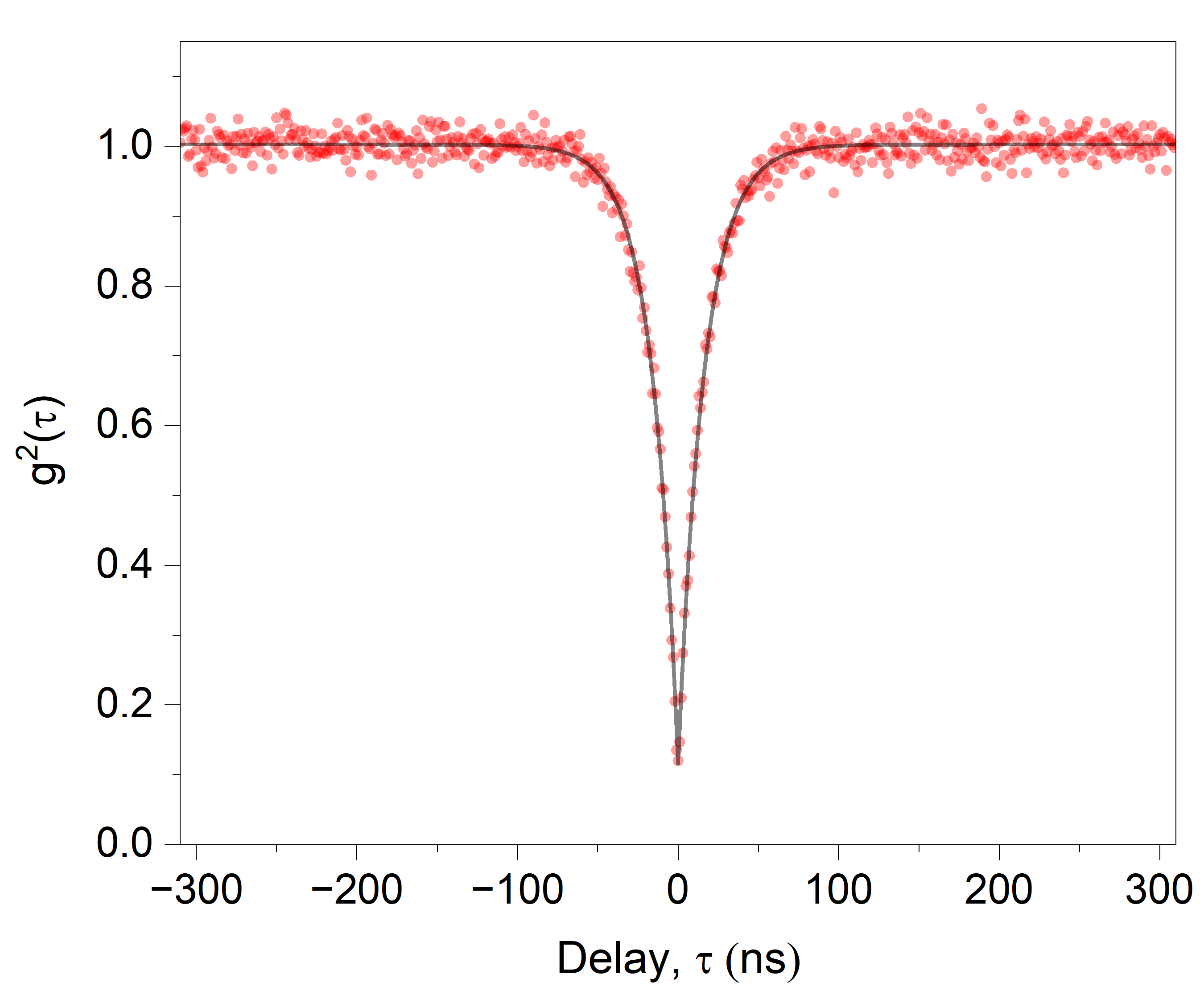}
 
\caption{Single-photon characteristics of the fiber-guided photons of the QRNG: The red dots represent the normalized second-order photon correlation function, $g^{(2)}(\tau)$ measured at an excitation intensity of 740 W/cm$^2$ , corresponding to the excitation intensity used to acquire the bits for the randomness analysis. The solid curve is the fit for the $g^{(2)}(\tau)$  with $g^{(2)}(0)$= 0.12$\pm$0.01.}
\label{Fig2}
\end{figure}

Each single-photon emission event results in one of two binary outcomes: a photon is detected at detector $D_1$ with a vacuum in $D_2$ or a photon is detected in $D_2$ with a vacuum in $D_1$. These states, $\ket{1}_{D_1} \ket{0}_{D_2}$ and $\ket{0}_{D_1} \ket{1}_{D_2}$ represent single-photon occupation in the propagating modes along the path of detector ($D_1$) and detector ($D_2$), respectively. For a spontaneous emission coupled to the ONF, the single-photon state may be modeled as:

\begin{equation}
\ket{\psi}
=
\frac{1}{\sqrt{2}}
\left(
\ket{1}_{D_1}\ket{0}_{D_2}
+
e^{i\phi}
\ket{0}_{D_1}\ket{1}_{D_2}
\right).
\label{eq1}
\end{equation}

where $\phi$ denotes the relative phase between propagation paths.

The quantum statistical properties of the emitted photons were characterized using the second-order correlation function $g^{(2)}(\tau)$, which quantifies intensity correlations between detection events at detectors $D_1$ and $D_2$. For our path-encoded detection scheme, the cross-correlation function is defined as:

\begin{equation}
g^{(2)}(\tau)
=
\frac{
\langle
\hat{a}_{D_1}^{\dagger}(t)
\hat{a}_{D_2}^{\dagger}(t+\tau)
\hat{a}_{D_2}(t+\tau)
\hat{a}_{D_1}(t)
\rangle
}{
\langle \hat{n}_{D_1}\rangle
\langle \hat{n}_{D_2}\rangle
}.
\label{eq2}
\end{equation}

where $\hat{a}_{D_1}$ and $\hat{a}_{D_2}$ are the annihilation operators for the optical modes incident on the detectors $D_1$ and $D_2$, respectively, and $\hat{n}_k\equiv\hat{a}_k^{\dagger}\hat{a}_k$ represents the photon number operator for each detection channel. The normalized correlation function provides crucial insights into the single-photon characteristics of the emitted photons.

Figure~\ref{Fig2} shows the normalized second-order photon correlation function, $g^{(2)}(\tau)$, of the QD emission in the system. The experimental data are represented by red dots, while the solid curve represents an exponential fit to the $g^{(2)}(\tau)$ data. The observed photon antibunching, characterized by $g^{(2)}(0)=0.12\pm0.01$, confirms the strong single-photon character of the QD emission. The measurement was performed at an excitation intensity of 740 W/cm$^2$, corresponding to the operating conditions used for random-bit generation.

\begin{figure}[!ht]

\centering

 \includegraphics[width=\columnwidth]{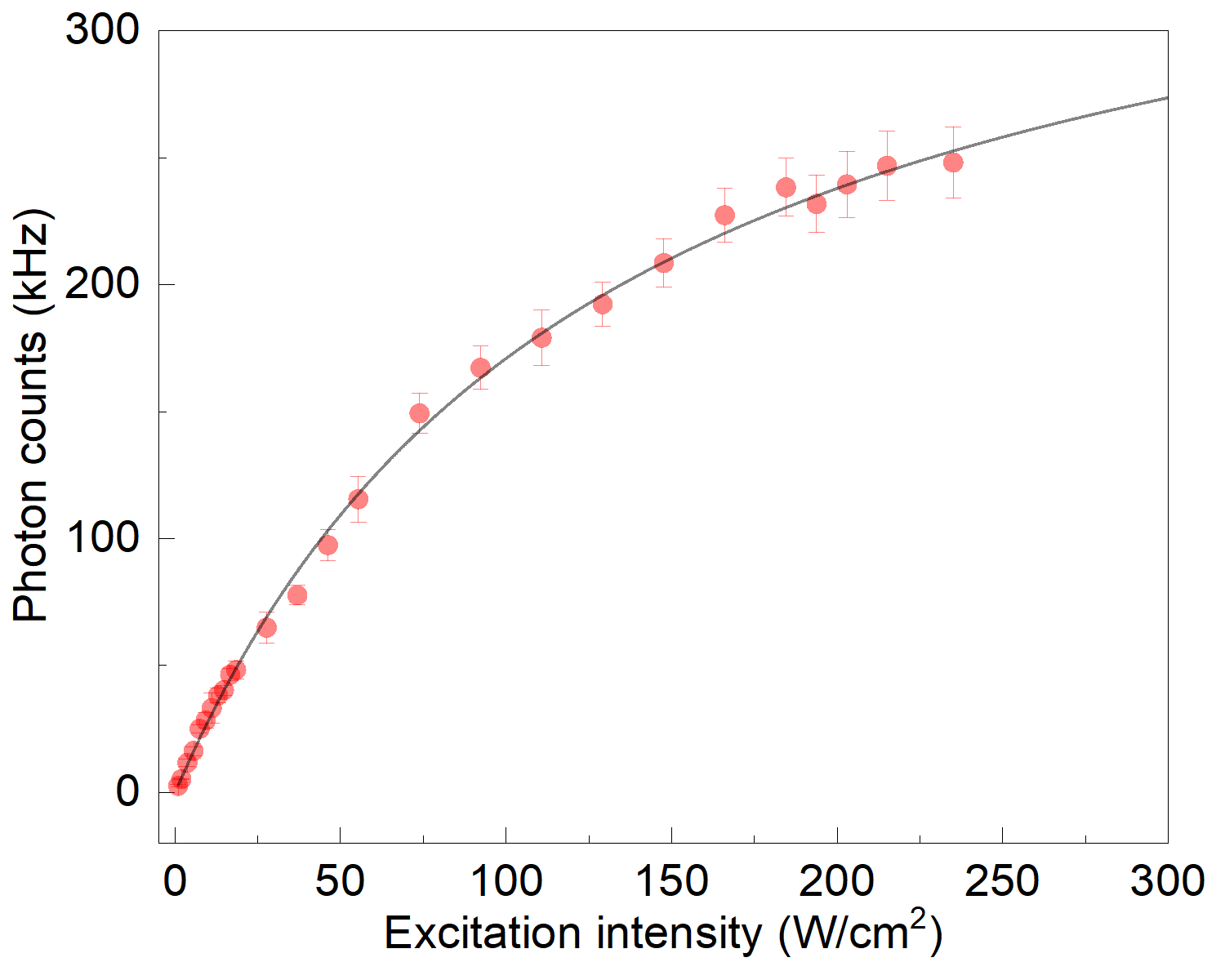}
 
\caption{ The red dots show the photon-count rate of single photons measured through one side of the fiber at different excitation intensity. The solid curve is the fit using the two-level saturation model given in Eq. \ref{eq5} with the saturated photon count rate $N_{\max}=0.371\pm0.017~\mathrm{MHz}$.} 
\label{Fig3}
\end{figure}

The maximum bit generation rate of the device was estimated from the measurement of photon-count rates, ($N(I)$) as a function of excitation intensity, $I$, and shown as red dots in Fig. \ref{Fig3}. 
The solid line represents the fit for the extracted photon-count rates using a two-level saturation model, 
\begin{equation}
{N(I)}= {N}_{max}\left[\frac{I}{I+I_{sat}}\right].
\label{eq5}
\end{equation}

Here, $N_{\max}=0.371\pm0.017$ MHz is the asymptotic saturated photon-count rate inferred from measurements at one side of the optical nanofiber, and $I_{\mathrm{sat}}=122\pm9$ W/cm$^2$ is the saturation intensity. The saturation intensity, $I_{\mathrm{sat}}$, corresponds to the excitation intensity at which the photon-count rate reaches half of its saturation value. The overall detection efficiency, $\alpha$, of the optical setup is estimated to be approximately 44\%, comprising the fiber-to-detector coupling efficiency in the presence of a 560~nm long-pass filter ($\sim74\%$) and the detector efficiency ($\sim60\%$). However, this detection efficiency is not used to correct the present bit-generation-rate calculation, since the QRNG bit rate is determined directly from the experimentally detected photon events. Because each detection event at either $D_1$ or $D_2$ generates one random bit, and the single photons are coupled symmetrically into the two counter-propagating guided modes of the ONF, the corresponding maximum raw bit-generation rate inferred from the saturation fit is
\begin{equation}
\Gamma_{\mathrm{QRNG}}
=2N_{\max}
=0.74\pm0.03~\mathrm{Mbps}.
\label{eq6}
\end{equation}
The factor of two accounts for the symmetric coupling of single photons into both propagating modes of the ONF. From the photon-count saturation measurements, we estimate an asymptotic maximum combined raw bit-generation rate of $0.74\pm0.03$ Mbps without any correction for optical transmission or detector efficiency.

\begin{figure}[!ht]

\centering

 \includegraphics[width=\columnwidth]{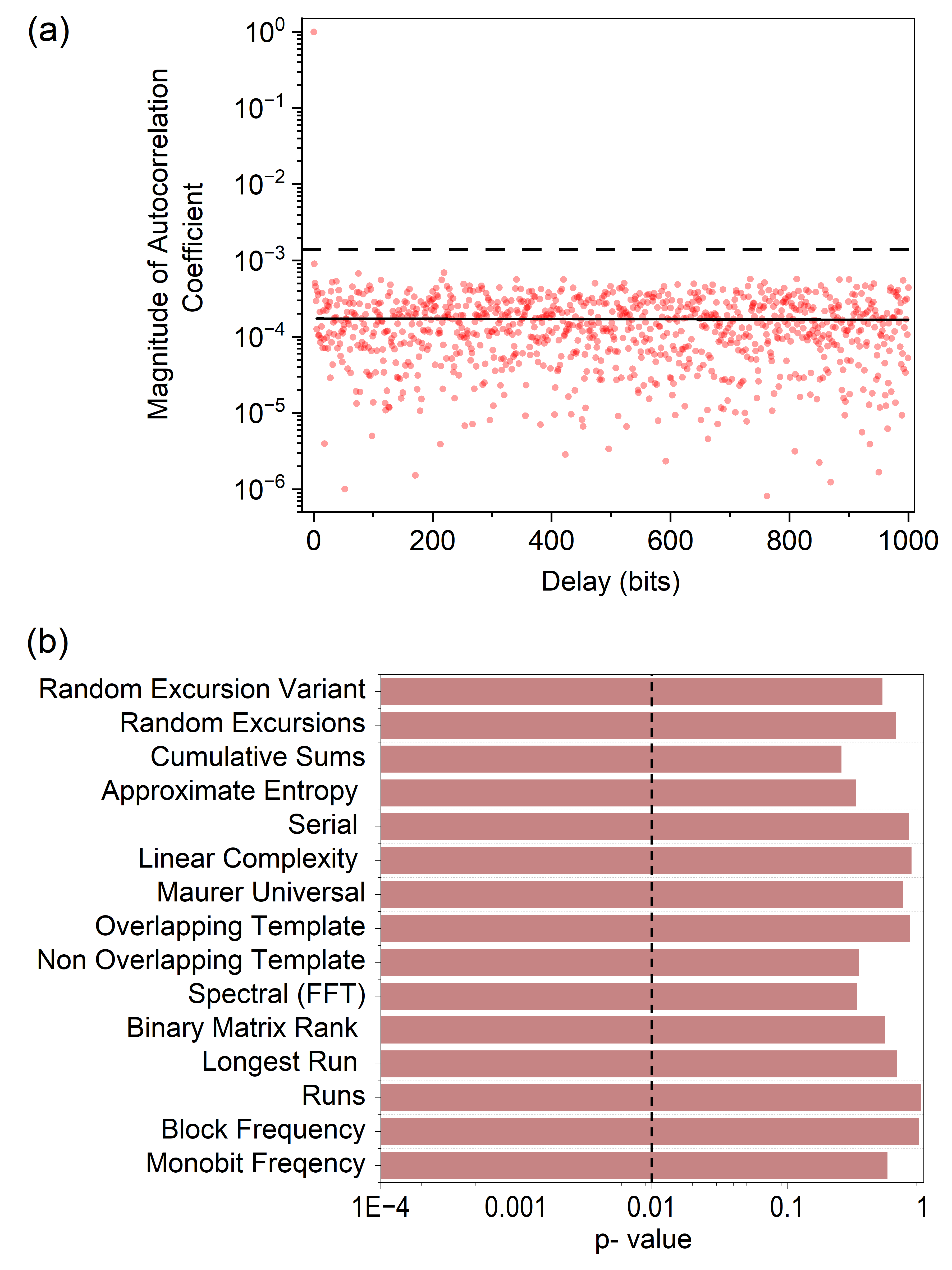}
 
\caption{An overview of the autocorrelation analysis and NIST Statistical Test Suite results for a raw bit sequence of 2 Mbits, generated from the photons with $g^{(2)}(0)$= 0.12$\pm$0.01. (a) Magnitude of autocorrelation coefficients of the bit stream. The solid line represents the mean value at 1.22$\times$10$^{-4}$ with a standard deviation of 1.02$\times$10$^{-4}$, while the dotted line indicates the 95$\%$ confidence limit. (b) NIST Statistical Test Suite results of 15 tests. The dotted line ($p$ = 0.01) represents the critical p-value above which the sequence is considered random.}
\label{Fig4}
\end{figure}

To assess the quality of randomness in the generated bitstream, autocorrelation coefficients were computed over a range of bit delays to identify any potential statistical dependencies. For a true random bit sequence, the autocorrelation coefficients at non-zero bit-delays fluctuate around zero within the statistical bounds of  $\pm2/\sqrt{N}$, where $N$ is the total number of bits in the sequence. This range, corresponding to the $95\%$ confidence interval, is a commonly accepted rule-of-thumb in hardware random number generator assessment and is consistent with the statistical acceptance criteria used for testing the independence of true random numbers  \cite{QCL10}. The randomness of the generated bit sequence were tested at various bit generation rates. The red dots in Fig.~\ref{Fig4}(a) show the magnitude of autocorrelation coefficients of a 2 Mbits sequence. The sequence was generated by exciting the single QD with an intensity of  740 W/cm$^2$.  At this operating intensity, the measured photon-count rate from each side of the fiber was 0.33 MHz, corresponding to a combined raw bit-generation rate of 0.66 Mbps.
The solid line at 1.22$\times$10$^{-4}$ represents the mean value of the autocorrelation coefficients, with a standard deviation of 1.02$\times$10$^{-4}$, which is significantly below the 95$\%$ confidence limit indicated by the dotted line. For better visualisation, only values up to a delay of 1000 bits are shown here.

The quality of randomness in the generated bit-stream was further evaluated and certified using the NIST Statistical Test Suite. The test suite determines the statistical probability (p-value) that the generated sequence is random. To ascertain the randomness of the generated raw bits, we carried out the 15 statistical tests. Fig.~\ref{Fig4}(b) shows an overview of the results from the raw bit sequence. The dotted line ($p$ = 0.01) represents the critical p-value above which the sequence is considered random. The bit sequence passed all 15 tests in the NIST suite. It is important to note that both the autocorrelation and NIST tests were performed without any post-processing of the raw bit stream, thereby highlighting the intrinsic quality of the generated random bits.

The fiber-integrated nanophotonic platform demonstrated here offers several practical advantages over conventional free-space or bulk optics QRNG implementations. The adiabatic tapering of the optical fiber ensures near-unity mode transformation efficiency, enabling an intrinsically alignment-free, compact, and mechanically stable device architecture under ambient room conditions. The tapered fiber plays a triple role: serving as the host platform for the emitter (here QD), acting as an efficient collector of the emitted single photons, and functioning as a low-loss waveguide for their propagation. The sub-wavelength dimensions of the host material substantially minimizes the background fluorescence from excitation, thereby increasing the purity of the detected quantum signal. This high intrinsic signal quality allows the raw bit sequence to be used directly for random number generation without the need for post-processing. In addition, the fiber-in-line configuration is inherently compatible with existing fiber-optic communication networks, paving the way for on-demand quantum randomness generation at the physical layer for quantum cryptography and secure key distribution protocols.

The quantum nature of the single-photon source was confirmed through second-order correlation measurements, yielding $g^{(2)}(0)$= 0.12$\pm$0.01. This low multi-photon probability, combined with the symmetric coupling of photons into counter-propagating modes of the ONF, produces an equiprobable path-encoded binary outcome that directly reflects the inherent quantum uncertainty of spontaneous emission. The statistical quality of the generated random numbers was validated through autocorrelation analysis and the NIST Statistical Test Suite. The raw bit sequences exhibited no significant autocorrelation, confirming statistical independence between successive bits. Furthermore, all 15 NIST tests were passed without any post-processing, underscoring the high statistical quality and robustness of the output, making it suitable for cryptographic applications where unpredictability and security are paramount. For completeness, the influence of multi-photon events on the QRNG performance has been systematically investigated and is discussed in the Supplemental Material.

The maximum detected bit-generation rate of the system is primarily limited by the brightness of the single-photon source. In the present ONF-based platform, the photon brightness is mainly constrained by the slow photoluminescence decay rate of the QD and the coupling efficiency of spontaneously emitted photons into the ONF-guided modes, which has a maximum limit of approximately 22\% for the present geometry. Although this performance is competitive with other state-of-the-art single-photon-based QRNGs \cite{LCF20,WKT20}, several strategies could enhance throughput. Integrating cavity structures on the tapered fiber could substantially improve coupling efficiency, with theoretical models predicting values up to 94\% for moderate-finesse cavities \cite{FamCavity, Yalla3, Schell}. Alternatively, a parallel fiber configuration \cite{TwinONF1} could achieve $>$50\% coupling efficiency according to numerical simulations. Such improvements would increase the number of detected single photons and consequently enhance the achievable random-bit generation rate while retaining the fiber in-line advantages of the present platform.

In conclusion, we have experimentally demonstrated a compact, alignment-free QRNG under ambient conditions, using single photons from a quantum dot coupled to an ONF. The ONF simultaneously hosts the emitter, collects photons, and guides them directly into a standard optical fiber, delivering high-purity single photons into counter-propagating guided modes of the fiber. The source generates pure single-photons with $g^{(2)}(0) = 0.12\pm0.01$. Together with symmetric coupling into counter-propagating fiber modes, the device yields high-quality quantum random bits and pass all tests of the NIST Statistical Test Suite. The device achieves a maximum asymptotic bit generation rate of $0.74\pm0.03~\mathrm{Mbps}$ with a very low autocorrelation coefficient.

By bridging nanoscale quantum emitters with existing macro-scale optical infrastructure, this fiber-integrated platform provides a promising route toward a deployable solution for on-demand quantum randomness generation in secure communication networks. With enhanced photon-collection architectures, such as cavity coupling or parallel-fiber geometries, the detected photon rate and consequently the random-bit generation rate could be substantially increased, paving the way toward scalable, high-speed quantum randomness generation for network-level applications.
\section*{Supplementary Material}
In the Supplementary Material, we provide additional experimental results investigating the effect of multi-photon events on the statistical quality of the generated random bits. The Supplemental Material compares the second-order photon correlation function and bitstream autocorrelation for single QD and two-QD photon sources, demonstrating the importance of the single-emitter regime for high-quality quantum random-number generation.

We acknowledge Kohzo Hakuta and C. M. Chandrashekhar for fruitful discussions. K.P.N. acknowledges support from a grant-in-aid for scientific research from the Japan Society for the Promotion of Science.

{}
\end{document}